\documentclass[nofootinbib,reprint,amsmath,amssymb,showkeys,superscriptaddress,showpacs,aps]{revtex4-1}

\usepackage[normalem]{ulem}
\usepackage{graphicx}
\usepackage{dcolumn}
\usepackage{xcolor,colortbl}
\usepackage{multirow}
\usepackage{bm}

\usepackage{hyperref}
\hypersetup{colorlinks, linkcolor={red},citecolor={blue},urlcolor={blue}}

\usepackage{accents}
\usepackage{scalerel}
\usepackage{tabularx,siunitx}
\usepackage{wasysym}
\usepackage{ mathrsfs }
\usepackage{comment}

\newcommand{\Lagr}{\mathcal{L}}

\newcommand{\ee}{\end{equation}}
\newcommand{\bea}{\begin{eqnarray}}
\newcommand{\eea}{\end{eqnarray}}
\def\ba{\begin{array}}
\def\ea{\end{array}}

\begin{document}

\title[Special issue EPJ plus]{Early and late time cosmology: the $f(R)$ gravity perspective}

\author{Francesco~Bajardi}
\email{francesco.bajardi@unina.it}
\affiliation{Scuola Superiore Meridionale,  Largo San Marcellino 10, 80138 Napoli, Italy}
\affiliation{Istituto Nazionale di Fisica Nucleare, Sezione di Napoli, Via Cinthia 80126 Napoli, Italy}

\author{Rocco~D'Agostino}
\email{rocco.dagostino@unina.it}
\affiliation{Scuola Superiore Meridionale,  Largo San Marcellino 10, 80138 Napoli, Italy}
\affiliation{Istituto Nazionale di Fisica Nucleare, Sezione di Napoli, Via Cinthia 80126 Napoli, Italy}

\author{Micol~Benetti}
\email{micol.benetti@unina.it}
\affiliation{Scuola Superiore Meridionale,  Largo San Marcellino 10, 80138 Napoli, Italy}
\affiliation{Istituto Nazionale di Fisica Nucleare, Sezione di Napoli, Via Cinthia 80126 Napoli, Italy}

\author{Vittorio~De~Falco}
\email{vittorio.defalco-ssm@unina.it}
\affiliation{Scuola Superiore Meridionale,  Largo San Marcellino 10, 80138 Napoli, Italy}
\affiliation{Istituto Nazionale di Fisica Nucleare, Sezione di Napoli, Via Cinthia 80126 Napoli, Italy}

\author{Salvatore~Capozziello}
\email{salvatore.capozziello@unina.it}
\affiliation{Scuola Superiore Meridionale,  Largo San Marcellino 10, 80138 Napoli, Italy}
\affiliation{Istituto Nazionale di Fisica Nucleare, Sezione di Napoli, Via Cinthia 80126 Napoli, Italy}
\affiliation{Dipartimento di Fisica ``E. Pancini", Universit\`a di Napoli ``Federico II", Via Cinthia 80126 Napoli, Italy}

\begin{abstract}
Discrepancies between observations at early and late cosmic epochs, and the vacuum energy problem associated with the  interpretation of  cosmological constant, are  questioning the  $\Lambda$CDM model. Motivated by these conceptual and observational facts, extensions of Einstein's gravity are recently intensively considered in view of curing unsolved issues suffered by General Relativity at ultraviolet and infrared scales. Here, we  provide a short overview of some aspects of $f(R)$ gravity,  focusing, in particular, on cosmological applications.  Specifically, Noether symmetries are adopted as a criterion to select viable models and investigate the corresponding dynamics. We thus find solutions to the cosmological field equations, analyzing the behaviour of  selected models from  the matter-dominated  to  the present epoch. Moreover,  constraints coming from  energy conditions and the so-called swampland criteria are also considered. In particular, we qualitatively discuss the possibility of $f(R)$ gravity to account for fixing cosmic tensions.
\end{abstract}


\maketitle
\section{Introduction}
\label{sec:intro}
More than one century after its formulation, General Relativity (GR) is continuously  confirmed as  a solid and self-consistent theory of gravity. Despite great successes, over the years,  several issues emerged  at  strong field regimes \cite{Joyce:2014kja, Ishak:2018his}, as well as at infrared scales \cite{Capozziello:2007ec}. Undoubtedly, GR perfectly works  at  Solar System scales, reproducing Newton's gravity in the weak field limit. As a matter of fact, GR predicts, with high precision, the perihelion precession, the Shapiro time delay, the light deflection and the gravitational lensing  at astrophysical scales \cite{Will:2014kxa, Salucci2020}. Nonetheless, GR fails in predicting  observations at different energy scales \cite{Clifton:2011jh}. For example, at infrared scales, a major problem is represented by the late-time cosmic speed-up, driven by a mysterious fluid with negative pressure, called ``dark energy", which should constitute the bulk  of the Universe content \cite{Riess1998,Perlmutter1999,Sahni1999,Peebles2003,Capozziello:2022jbw}. In this respect,  GR is not capable of explaining acceleration effects, so one needs to include the cosmological constant $\Lambda$ in the gravitational action. Albeit $\Lambda$ gives \emph{de facto} the simplest interpretation for accelerating expansion,  it suffers from theoretical shortcomings, such as the huge discrepancy between the value coming from observations and the vacuum energy density inferred from quantum field theory \cite{Weinberg1989,Padmanabhan2003,DAgostino:2022fcx}. From the observational point of view, recent data highlighted cosmological issues questioning the standard $\Lambda$CDM model, among the others the $H_0$ and $\sigma_8$ tensions \cite{Lukovic:2016ldd,Verde:2019ivm}. In particular, discrepancies between the direct (model-independent) measurement of the Hubble parameter by Cepheids and other standard candles and its value inferred from Planck satellite data \cite{Planck:2018jri} represent a challenging puzzle in our understanding of the cosmic expansion \cite{Riess:2021jrx}. All these arguments motivate to explore alternatives to GR, ranging from dynamical scalar fields  \cite{Ratra1988,Armendariz-Picon2000,Copeland2006, SantosDaCosta:2018bbw}, extensions of GR \cite{Capozziello2011,Nojiri2010}, unified models accounting for dark matter and dark energy \cite{Scherrer2004,Capozziello:2017buj,Capozziello:2018mds,Brandenberger2019,DAgostino:2021vvv, Benetti:2019lxu, Benetti:2021div}, up to scenarios based on the holographic principle \cite{Li2004,DAgostino:2019wko,Saridakis2020}. In particular, the dark matter issue can be considered under the standard of extended theories of gravity  \cite{Capozziello:2012ie, Napolitano:2012fp}.

Moreover, the formalism of GR seems to be incompatible with other fundamental interactions, mainly due to the impossibility of defining a Hilbert space in a formal way. This renders GR unviable from a quantum point of view. Therefore, no self-consistent theory of quantum gravity is so far capable of addressing all the high-energy shortcomings of Einstein's theory. In addition, ultraviolet divergences, arising when expanding the Hilbert-Einstein action up to the second-loop level, cannot be canceled by renormalization procedures \cite{Goroff1985}. To the purpose of addressing all these (and several other) issues, gravitational alternatives to GR may be considered \cite{Capozziello2011,Nojiri2017,Capozziello2019}. Some of them extend the Hilbert-Einstein action by including a function of the scalar curvature, $R$, giving rise to the so-called $f(R)$ gravity theories \cite{Capozziello:2002rd,Sotiriou2008,DeFelice2010,Nojiri2010,Capozziello:2018aba,DAgostino:2020dhv}. However, the gravitational action can be generalized in several ways, such as introducing  couplings between geometry and dynamical scalar fields \cite{Fujii:2003pa, Abedi2018, DAgostino:2018ngy, Bajardi:2020xfj}, higher-order derivatives \cite{Wands:1993uu, Capozziello:2021krv, Gottlober:1989ww}, or further curvature invariants \cite{Bajardi:2020osh, Nojiri:2005jg, Wheeler:1985nh, Bajardi:2022tzn}.
Other possibilities include also relaxing Lorentz invariance principle \cite{Horava:2009uw, Sotiriou:2010wn, Carroll:2001ws}, or considering dynamics ruled by torsion \cite{Bajardi:2021tul, Capozziello:2022vyd, Linder:2010py, Bengochea:2008gz, Cai:2015emx} and non-metricity \cite{Capozziello:2022wgl, Bajardi:2020fxh, BeltranJimenez:2019tme, BeltranJimenez:2017tkd}.

In this paper, we focus on the $f(R)$ extension of GR, whose field equations, in metric formalism, are of the fourth order. It is interesting to point out that the right-hand side of  $f(R)$ field equations can be understood as an effective stress-energy tensor generated by geometry, capable, in principle, of mimicking dark energy without introducing, by hand, the cosmological constant $\Lambda$. In this regard, it can be shown that the issues of quintessence and cosmic acceleration can be framed within some $f(R)$ gravity models \cite{Capozziello:2002rd,Hu:2007nk, Tsujikawa:2007xu,Capozziello:2022rac}. For some selected functions, also the galaxy rotation curve can be fitted, without resorting to dark matter \cite{Capozziello:2012ie, Capozziello:2006ph, Famaey:2011kh, Boehmer:2007kx}. An important  $f(R)$ extension is the Starobinsky model \cite{Starobinsky:1980te}, where the action contains an extra quadratic term of the form $\alpha R^2$. This theory gained great success due to the excellent match between theoretical predictions and observational data for describing  the dynamical evolution of the Universe during the early stages \cite{Planck:2018jri}. In this framework,  the quadratic term in the Ricci scalar $R$ is dominant only in the early times and subsequently slowly decreases along the Universe evolution.

It is worth  stressing that no $f(R)$ model (or, in general, no alternative to GR) seems to fit  the whole set of cosmic observations at once, fixing simultaneously all the incompatibilities at the theoretical level. As mentioned above, some $f(R)$ models can settle part of such shortcomings, but although they may turn out to be valid at some scales, they are often ruled out by experiments at different energies, where other modifications result to be more appropriate \cite{deMartino:2020yhq}. The goal of a  unified theory of gravity, valid at all scales, could be very difficult to achieve. In this regard, one can assume the gravitational action to be made of different terms, with corresponding coupling constants becoming dominant at certain scales \cite{DeLaurentis:2015fea}. In other words, the action could be made of different contributions, which become dominant or  subdominant at specific scales. These remarks show that a model constituted by piecewise modified gravity theories, acting at the appropriate scales, seems to be actually  one of the feasible solutions to deal with the whole cosmological dynamics \cite{Capozziello:2010zz}.

The purpose of this paper is to point out some cosmological applications of $f(R)$ gravity at early and late times, in view of possible solutions to some issues of modern cosmology. It is organized as follows. 
In Sec.~\ref{NoethSec}, $f(R)$ cosmological models are selected by means of the Noether symmetry approach. In Sec.~\ref{studyselectedmodel}, we investigate models in a Friedman-Lama\^itre-Robertson-Walker (FLRW)  background, analyzing the corresponding equation of motion. In Sec.~\ref{CosmoSec},  the cosmological behaviour of the selected $f(R)$ models is studied and  theoretical bounds over the free parameters are outlined.  
In Sec.~\ref{ECSR}, we consider  energy conditions while   swampland criteria are taken into account in Sec.~\ref{SwampSec}. In Sec.~\ref{Concl},  final considerations are reported and future perspectives are outlined. 
Throughout the paper, we use units  $8 \pi G = c = \hbar=1$, unless otherwise indicated.

\section{Selecting $f(R)$ models}
\label{NoethSec}
Being $f(R)$ a  general function of the Ricci scalar $R$, one needs a physical criterion to select viable models. A possible approach can be based on the existence of Noether symmetries for $f(R)$ dynamics \cite{Capozziello:1996bi,Capozziello:2008ch}.
Here, we introduce the $f(R)$ gravity field equations and select viable models according to the existence of Noether symmetries. Specifically, the $f(R)$ action reads 
\begin{equation}
    S = \int d^4x\, \sqrt{-g} \left[ \frac{1}{2}f(R) +\mathcal{L}_m\right],
\end{equation}
being $g$ the determinant of the metric tensor $g^{\mu\nu}$ and $\mathcal{L}_m$ the matter Lagrangian density. Varying the above action with respect to $g_{\mu\nu}$, one obtains the field equations
\begin{equation}
f_R \, G_{\mu \nu} =\frac{1}{2} g_{\mu \nu} \left[f - R f_R \right] + f_{R{; \mu ; \nu}} - g_{\mu \nu} \Box f_R + T_{\mu \nu}\,,
\label{fematterfR}
\end{equation}
where $f_R \equiv \frac{\partial f}{\partial R}$, $G_{\mu \nu}\equiv  R_{\mu \nu} - \frac{1}{2} g_{\mu \nu} R$ is the Einstein tensor and the semicolon denotes the covariant derivative. Here, $T_{\mu\nu}$ is the energy-momentum tensor of matter fields,
\begin{equation}
T_{\mu\nu}=-\dfrac{2}{\sqrt{-g}}\dfrac{\delta \mathcal{L}_m}{\delta g^{\mu\nu}}\,,
\end{equation}
satisfying the conservation law $T_{\mu\nu}{}^{;\nu}=0$. It is worth to note that, for $f=R$, \emph{i.e.} $f_R = 1$, Einstein's field equations are recovered. 

Among the  possible $f(R)$ extensions of GR, we can single out models containing symmetries, based on Noether's theorem. Symmetries, in fact, allow us to reduce the minisuperspace dimension and find analytic solutions for the given dynamical system \cite{Capozziello:2008ch, Urban:2020lfk, Dialektopoulos:2018qoe}. The Noether Symmetry Approach consists of assuming that there exists a point transformation leaving the point-like Lagrangian invariant, whose generator $X$ depends on the minisuperspace variables. For instance, in a cosmological and spatially flat space-time, the Noether vector reads 
\begin{equation}
X = \eta^i \frac{\partial}{\partial q^i} + \dot{\eta}^i \frac{\partial}{\partial \dot{q}^i},
\end{equation}
with $\eta^i$ being unknown functions of the minisuperspace variables $q^i$. The first prolongation of $X$, including the first derivative transformation, is
\begin{equation}
    X^{[1]} = \xi \frac{\partial}{\partial t} + \eta^i \frac{\partial}{\partial q^i} + \left(\dot{\eta}^i - \dot{q}^i \dot{\xi}\right) \frac{\partial}{\partial \dot{q}^i}.
\end{equation}
According to Noether's theorem, if the following condition holds,
\begin{equation}
    X^{[1]} \Lagr + \dot{\xi} \Lagr = \dot{\gamma},
    \label{noethidentity}
\end{equation}
then $X$ is a symmetry generator and the quantity
\begin{equation}
   I = \left(\eta^i-\xi \dot{q}^i \right) \frac{\partial \Lagr}{\partial \dot{q}^i} + \xi \Lagr - \gamma,
   \label{consquant}
\end{equation}
is an integral of motion. In Eqs.~\eqref{noethidentity} and \eqref{consquant}, $\Lagr$ is the point-like Lagrangian, $\xi$ is the infinitesimal generator of time transformations and $\gamma$ is a gauge function of $t$ and $q^i$. Therefore, in $f(R)$ cosmology, where the minisuperspace is $\mathcal{S} = \{a,R\}$ (with $a$ being the cosmological scale factor), the first prolongation of the Noether vector reads
\begin{equation}
     X^{[1]} = \xi \frac{\partial}{\partial t} + \alpha \frac{\partial}{\partial a} + \beta \frac{\partial}{\partial R} +  \left(\dot{\alpha} - \dot{a} \dot{\xi}\right) \frac{\partial}{\partial \dot{a}} + \left(\dot{\beta} - \dot{R} \dot{\xi}\right) \frac{\partial}{\partial \dot{R}},
\end{equation}
where $\xi$, $\alpha$ and $\beta$ are functions of $t, a$ and $R$. 

In order to find the point-like Lagrangian with respect to a spatially flat FLRW metric of the form $ds^2 = dt^2 - a(t)^2 d$\textbf{x}$^2$, let us consider the $f(R)$ action in vacuum, written in terms of Lagrange multipliers $\lambda$, namely
\begin{equation}
    S = \int dt \left[a^3 f(R) - \lambda \left(R + 6 \frac{\ddot{a}}{a} + \frac{\dot{a}^2}{a^2} \right)\right].
    \label{actionf(R)}
\end{equation}
We assume the Lagrangian to be depending only on the cosmic time through the scale factor and the Ricci scalar. Under this assumption, we can integrate the three-dimensional surface. By varying the action with respect to $\lambda$ and integrating out higher derivatives, the point-like cosmological Lagrangian turns out to be
\begin{equation}
    \Lagr =  a^3 \left[ f - R f_R \right] + 6 a \dot{a}^2 f_R + 6 a^2 \dot{a} \dot{R} f_{RR}\,.
    \label{Lagrf(R)}
\end{equation}
Dynamics is given either by the $f(R)$ gravity field equations or by the Euler-Lagrange equations along with the energy condition $E_{\Lagr} = \dot{q}^i \partial \Lagr/\partial \dot{q}^i - \Lagr = 0$. Both approaches provide a set of three differential equations of the form
\begin{subequations}
\begin{align}
&6 a^2 \dot{a} \dot{R} f_{RR} + 6 a \dot{a}^2 f_R - a^3[f    - R f_R] = 0\,,
\\
&\dot{R}^2 f_{RRR}  + \ddot{R} f_{RR} + \frac{\dot{a}^2}{a^2} f_R + 2 \frac{\ddot{a}}{a} f_R\,,
\\
& \frac{1}{2} [f - R f_R] - 2\frac{\dot{a}}{a} \dot{R} f_{RR} = 0\,,
\\
&R = - 6 \left( \frac{\ddot{a}}{a}+  \frac{\dot{a}^2}{a^2} \right). \label{ELR}
\end{align}
\end{subequations}
Notice that the equation of motion with respect to $R$, namely Eq.~\eqref{ELR}, provides the cosmological expression of the scalar curvature, as expected by construction. Clearly, the system can be solved only after selecting the corresponding model. To this purpose, the application of the Noether Symmetry Approach to Lagrangian \eqref{Lagrf(R)} yields a system of 9 partial differential equations, not all independent \cite{Capozziello:2008ch, Fazlollahi:2018wmp}. Decomposing $\eta^i$ as $\eta^i \to \{\alpha, \beta\}$, the Noether symmetry existence condition allows to select 8 models with symmetries. The list of functions with the related infinitesimal generators is reported in Table \ref{Noether solutions f(R)}.
\begin{table}
    \centering
    \caption{Noether solutions in $f(R)$ cosmology.} 
\begin{tabular}{l c c c c}\hline\hline 
\multicolumn{1}{c}{$\xi(t)$} & $\alpha(a)$ & $\beta(a,R)$ & $f(R)$
\\  \hline \\ 
$ \xi_0 t + \xi_1 \,\,\,\,$& $\displaystyle \sqrt{\alpha_0 a + \alpha_1}\,\,\, $ & $ \beta(a,R,t)\,\,\, $ &$ f_0 R + \Lambda $  \\ \\
$ \xi_0 t + \xi_1 $& $ \displaystyle \frac{7 \xi_0}{36} a $ & $\displaystyle  -\frac{7 \xi_0}{6} R  $ & $  \displaystyle f_0 R^{\frac{19}{14}} + \Lambda $  \\  \\
$  \xi_0 t + \xi_1 $ & $ \alpha_0 a $ & $ - 2 \xi_0 R $& $ f_0 R^{1 + \frac{3 \alpha_0}{2 \xi_0}} $ \\ \\
$ \xi_0 $& $ 0 $ & $ 0 $ & $ \displaystyle  f_0 R^{\frac{19}{14}} + \Lambda $  \\  \\
$ 0 $& $ \displaystyle \frac{\alpha_0}{a^2} $ & $\displaystyle - 3 \alpha_0 \frac{R}{a^3}  $ & $  f_0 R  $  \\  \\
$ 0 $& $ \displaystyle \alpha_0 \sqrt{a} $ & $\displaystyle -  \frac{\alpha_0}{2} \frac{R}{\sqrt{a^3}}  $ & $   f_0 R $  \\  \\
$ 0 $& $ \displaystyle \frac{\alpha_0}{a} $ & $ \displaystyle -  2 \alpha_0 \frac{R}{a^2}  $ & $  \displaystyle f_0 R^{\frac{3}{2}}$  \\  \\
$ 0 $& $ 0 $ & $\displaystyle \frac{\beta_0}{a}  $ & $\displaystyle   f_0 R + f_1 R^2 $  \\  \\
\hline \hline 
\end{tabular}
\label{Noether solutions f(R)}
\end{table}
In the following, we focus our attention on the third solution, which is interesting for cosmological purposes. Specifically, the quantity $\frac{3 \alpha_0}{2 \xi_0}$ can measure deviations from GR so that, when $\alpha_0 = 0$, Einstein's theory is fully recovered. As a matter of fact, setting $\epsilon \equiv \frac{3 \alpha_0}{2 \xi_0}$, the gravitational action can be written as
\begin{equation}
S =\frac{1}{2} \int \sqrt{-g} \, R^{1+\epsilon} \, d^4 x\,,
\end{equation}
and, moreover, when $\epsilon \ll 1$ the function can be expanded up to the second order, providing
\begin{equation}
f(R)\approx R+ \epsilon R \log R + \mathcal{O}(\epsilon^{2}).
    \label{eq:powerlawf1}
\end{equation}
In order not to lose generality, in what follows we consider the general case with an arbitrary exponent, namely we set $k \equiv \frac{3 \alpha_0}{2 \xi_0} +1$.

\section{The $f(R)=R^{k}$ model}
\label{studyselectedmodel}
In view of the aforementioned considerations and taking into account the third solution of Table \ref{Noether solutions f(R)}, Eq.~\eqref{eq:powerlawf1} becomes
\begin{equation}
f(R)\approx R+ (k-1) R \log R + \mathcal{O}(k^{2}),
    \label{eq:powerlawf}
\end{equation}
where $k = 1 + \epsilon$. Such a function has been considered in the literature in different contexts. The constant $k$ can be thought of as a controlling parameter that quantifies deviations from the Hilbert-Einstein action. In the limit $|k-1|\ll 1$, namely when the model slightly deviates from GR, $f(R)$ as in Eq.~\eqref{eq:powerlawf} can be expanded up to the second order. 
%

This model exhibits many interesting features at different scales. As mentioned above, under given limits, it can fit the galaxy rotation curve without any dark matter \cite{Capozziello:2012ie}, while, at cosmological scales, quintessence can be addressed by further geometric contributions  stemming from the field equations \cite{Capozziello:2002rd}. Furthermore, the inflationary epochs can be predicted without additional scalar fields \cite{Nojiri:2017ncd}. Similarly to  other $f(R)$ models, a perturbation of the metric around the flat space-time yields gravitational waves with additional polarization modes \cite{Capozziello:2019klx}. In this regard, the production of gravitational waves can be achieved by this model in the early Universe \cite{Capozziello:2007vd}, as well as small deviations by the apsidal motion of eccentric binary stars \cite{DeLaurentis:2012dq}. Furthermore, this kind of models has been considered to study null and time-like geodesics in the weak field limit, with applications to the Solar System \cite{Clifton:2005aj} and black holes  \cite{Capozziello:2007id}. 

Clearly, no $f(R)$ theory so far is capable of tracing the whole history of the Universe and this is the reason why scenarios like $R^{k}$ are often considered just as \emph{toy models}. However, on the one hand, they can account for effective models, providing hints for a yet unknown  unified theory. On the other hand, they can indicate to what extent deviations from GR can affect the dynamics of the system.

It is also interesting to notice that any fourth-order theory of gravity, like the model considered here, under conformal transformations is dynamically equivalent to a second-order theory non-minimally coupled with a dynamical scalar field. As a matter of fact, through the conformal rescaling of the metric $g_{\mu \nu} \to \tilde{g}_{\mu \nu} = e^{2 \omega} g_{\mu \nu}$, the Levi-Civita connection, the Ricci tensor and the Ricci scalar can be recast in the Einstein frame as, respectively,
\begin{align}
&\tilde{\Gamma}^\sigma_{\lambda \mu} = \hat{\Gamma}^\sigma_{\lambda \mu} + g^{\nu \sigma} \left(\partial_\lambda \omega \; g_{\mu \nu} + \partial_\mu \omega \; g_{\lambda \nu} - \partial_\nu \omega \; g_{\lambda \mu} \right), \\
&\tilde{R}_{\alpha \beta} = R_{\alpha \beta} - 2 \omega_{; \alpha ; \beta} + 2 \omega_{; \alpha} \omega_{; \beta} - g_{\alpha \beta} \Box \omega - 2 g_{\alpha \beta} \; \omega_{; \gamma} \omega^{; \gamma}\,, \\
&\tilde{R} = e^{-2 \omega} \left(R - 6\Box  \omega - 6 \omega_{; \gamma} \omega^{; \gamma} \right).
\label{gamma conforme}
\end{align}
One can then define $\omega = \frac{1}{2} \ln |f_R(R)|$ and consider the conformal rescaling factor $\omega = \sqrt{\frac{1}{6}} \phi$, so that the field equations take a GR-like form, \emph{i.e.} $\tilde{G}_{\alpha \beta} = \tilde{T}^{\,\phi}_{\alpha \beta}$, where 
\begin{align}
\label{Fluidfi}
&\tilde{T}^{\,\phi}_{\alpha \beta} = \phi_{; \alpha} \phi_{; \beta} - \frac{1}{2} \tilde{g}_{\alpha \beta} \phi_{; \gamma} \phi^{; \gamma} + \tilde{g}_{\alpha \beta} V(\phi),
\\
&V(\phi) = \frac{1}{2} \frac{f(R) - R f_R}{[f_R]^2}.
\label{potentialconf}
\end{align}
In our case, considering a conformal transformation of Eq.~(\ref{eq:powerlawf}), one obtains a scalar-tensor model with a potential of the form
\begin{equation}
\label{eq:V_powerlaw}
V(\phi) = \frac{k-1}{2k^2} \left( \frac{e^{2\phi}}{k} \right) ^{\frac{k}{k-1}}.
\end{equation}
From Eq.~\eqref{potentialconf}, it is possible to notice that the cosmological constant is recovered when  $k \rightarrow 2$. This is due to the fact that $R^2$ is conformally invariant and behaves like a constant in the gravitational field equations. Also, Eq.~\eqref{eq:V_powerlaw} suggests that, when $k \rightarrow \infty$, it goes to an exponentially suppressed plateau. Both values of $k$ indicate that different models intrinsically contain the cosmological constant as a natural feature, at least asymptotically. Specifically, in the first limit, the potential in Eq.~\eqref{eq:V_powerlaw} turns out to be trivially constant, leading therefore to GR non-minimally coupled to a scalar field in presence of a cosmological constant. In the second limit, the potential takes the form $V(\phi) = e^{-2 \phi}/2$ and asymptotically converges to $\Lambda$, so that the resulting theory is a second-order scalar-tensor model with a  cosmological constant. As a consequence, the first case can be used to investigate dynamics at infrared regimes, while the second at ultraviolet ones.

In \cite{Stewart:1994ts, Lyth:1998xn, Kallosh:2013yoa}, the authors argue that models of this form may be naturally obtained even if the original potential is not particularly flat, when the early stages of the Universe are considered. 

Let us now focus on the cosmological solutions of $f(R)\sim R^{k}$ gravity in vacuum, which will be used in the next sections in order to constrain the free parameters of the $R^{k}$ model. 
Plugging the selected function into Eq.~\eqref{Lagrf(R)}, we have
\begin{equation}
\Lagr = f_0 a R^{k} \left(6 k (k-1) a \dot{a} \dot{R} +6 k R \dot{a}^2- (k-1) a^2 R^2\right).
\end{equation}
The system of equations of motion can be now solved analytically, providing
\begin{equation}
a(t) = a_0 t^{\frac{(k-1)(2k-1)}{2-k}}, \quad R(t) = \frac{6 k (k-1) (5-4k) (2k-1)}{(k-2)^2 t^2},
\label{solf(R)Rk}
\end{equation}
which hold only for $ k \neq 2$. When $k = 2$, exponential solutions occur:
\begin{equation}
a(t) = a_0 e^{\ell t}, \quad R(t) = -12 \ell^2, \quad k \to 2,
\label{expsolf(R)}
\end{equation}
with $\ell$ being a real constant. Finally, an interesting case is given by $k = 3/2$, that is the so-called \emph{Liouville field theory} and it is one of the few cases where a fourth-order Lagrangian can be expressed (in the Einstein frame) in terms of elementary functions under a conformal transformation \cite{Capozziello:1996xg}. In such a case, we obtain another class of solutions of the form
\begin{equation}
a(t)=a_0 \left[c_{4} t^{4}+c_{3} t^{3}+c_{2} t^{2}+c_{1} t+c_{0}\right]^{1 / 2},
\label{cosmosolf(R)}
\end{equation}
where $c_i$ are integration constants that can be fixed through observations at astrophysical and cosmological levels. Here, with the purpose to verify whether geometrical contributions are capable of mimicking the cosmological constant, we have not included matter Lagrangians to solve the cosmological field equations. As a result, we found that exponential and power-law scale factors are solutions of the  Euler-Lagrange equations.   In what follows, we investigate the cosmological behaviour of the $f(R) \sim R^k$ model in presence of matter fields.

\section{Cosmological dynamics}
\label{CosmoSec}

In this section, we analyze the cosmological dynamics of $f(R)$ theories in the metric formalism.

To obtain the cosmological solutions to the field equations,  we use the previously considered spatially-flat FLRW metric with the cosmic scale factor conventionally normalized to the unity at the present time, i.e. $a(t_0)\equiv a_0=1$. The Ricci scalar can be thus written in terms of the Hubble parameter $H\equiv \dot{a}/a$ as
\begin{equation}
R=-6(2H^2 +\dot H)\,,
\end{equation}
where the dot denotes the time derivative. Assuming that the Universe is filled with a perfect fluid, we have
\begin{equation}
{T^{\mu}}_{\nu}^{(m)}=\text{diag}(\rho_m,\, - p_m, \, - p_m, \, - p_m)\,,
\end{equation}
where $\rho_m$ and $p_m$ are the matter density and pressure, respectively.
Neglecting the late-time contribution of radiation and assuming non-relativistic pressureless matter obeying the continuity equation
\begin{equation}
\dot{\rho}_m+3H\rho_m=0\,,
\end{equation}
we can write the modified Friedman equations as
\begin{align}
3f_R H^2 & =\frac{1}{2} (f_R R-f)-3H\dot{f}_R+ \rho_m\,, \label{rhoGF} \\
-2f_R\dot{H} & =  \ddot{f}_R -H \dot{f} + \rho_m\,.
\label{pressGF}
\end{align}

One may study the cosmic dynamics by introducing the following dimensionless quantities \cite{Amendola:2006eh}:
\begin{equation}
x_1\equiv -\dfrac{\dot{f}_R}{H f_R}\,, \quad x_2\equiv -\dfrac{f}{6f_R H^2}\,, \quad \Omega_m\equiv \dfrac{\rho_m}{3f_RH^2}\,.
\end{equation}
For the specific model $f(R)=f_0 R^k$, where $\alpha$ and $k$ are constant parameters, the cosmological equations can be recast into the following system of coupled differential equations:
\begin{subequations}
\begin{align}
& x_1'=x_1^2+k x_2(x_1+1)-3x_2-1 \,, \label{eq:x1}\\
& x_2'=\frac{k}{1-k}x_1x_2+x_2(x_1+2k x_2+4)\,. \label{eq:x2}
\end{align}
\end{subequations}
Here, the prime denotes the derivative with respect to $N\equiv \ln a$. We also have the constraint equation
\begin{equation}
\Omega_m=1-x_1+x_2(k-1)\,.
\end{equation}

Theoretical bounds over $k$ may be obtained from the effective Equation of State (EoS) parameter, given by
\begin{equation}
w_\text{eff}\equiv -1- \dfrac{2}{3}\dfrac{\dot H}{H^2}\,.
\label{eq:weff}
\end{equation}
In fact, for the $f(R)$ model under consideration, we have
\begin{equation}
x_2=-\dfrac{1}{k}\left(2+\dfrac{\dot H}{H^2}\right),
\end{equation}
which allows to write
\begin{equation}
w_\text{eff}=\dfrac{1}{3} (1+2 k x_2)\,.
\label{eq:weff_n}
\end{equation}

Since any viable $f(R)$ model must approach the $\Lambda$CDM paradigm for $a\ll 1$, we consider the expansion history in the matter-dominated (MD) era given by
\begin{equation}
H(a)=H_0\sqrt{\Omega_{m0}\, a^{-3}+1-\Omega_{m0}}\,,
\end{equation}
which implies $w_\text{eff}^\text{(MD)}\simeq 0$.
The $\Lambda$CDM model corresponds to $f(R)=R-2\Lambda$, where $\Lambda=3H_0^2(1-\Omega_{m0})$, and $f_R=1$. We thus obtain
\begin{equation}
x_2=\dfrac{2 a^3 (1-\Omega_{m0})+\Omega_{m0}}{2 a^3 (\Omega_{m0}-1)-2\Omega_{m0}} \xrightarrow[a\ll 1]{}  -\dfrac{1}{2}\,,
\label{eq:x2_n}
\end{equation}
and, from Eq.~\eqref{eq:weff_n}, we find
\begin{equation}
w_\text{eff}^\text{(MD)}\simeq \dfrac{1-k}{3}\,.
\end{equation}
Therefore, the considered $f(R)$ model is able to reproduce the MD era only for $|k-1|\ll 1$.
To determine the cosmological behaviour up to the present time, we need to solve the system \eqref{eq:x1}--\eqref{eq:x2}. This can be done numerically by means of suitable initial conditions over $(x_1,x_2)$ calculated in the MD era. Once $x_2$ is known, one can then solve Eq.~\eqref{eq:x2_n} to obtain $H(a)$. Therefore, a direct comparison with observations can provide information on the free parameter(s) of the $f(R)$ model.

\subsection{Curvature Quintessence}
Let us now turn our attention to $f(R)$ cosmology in vacuum, assuming that at very early and late times matter contribution is subdominant with respect to geometrical effects. In this regards, the cosmic acceleration can be accounted for by means of curvature invariants, without the need for the cosmological constant. The resulting scenario is referred to as \emph{curvature quintessence}  \cite{Capozziello:2002rd, Capozziello:2003gx}. 

\begin{figure}
\centering
\includegraphics[width=3.3in]{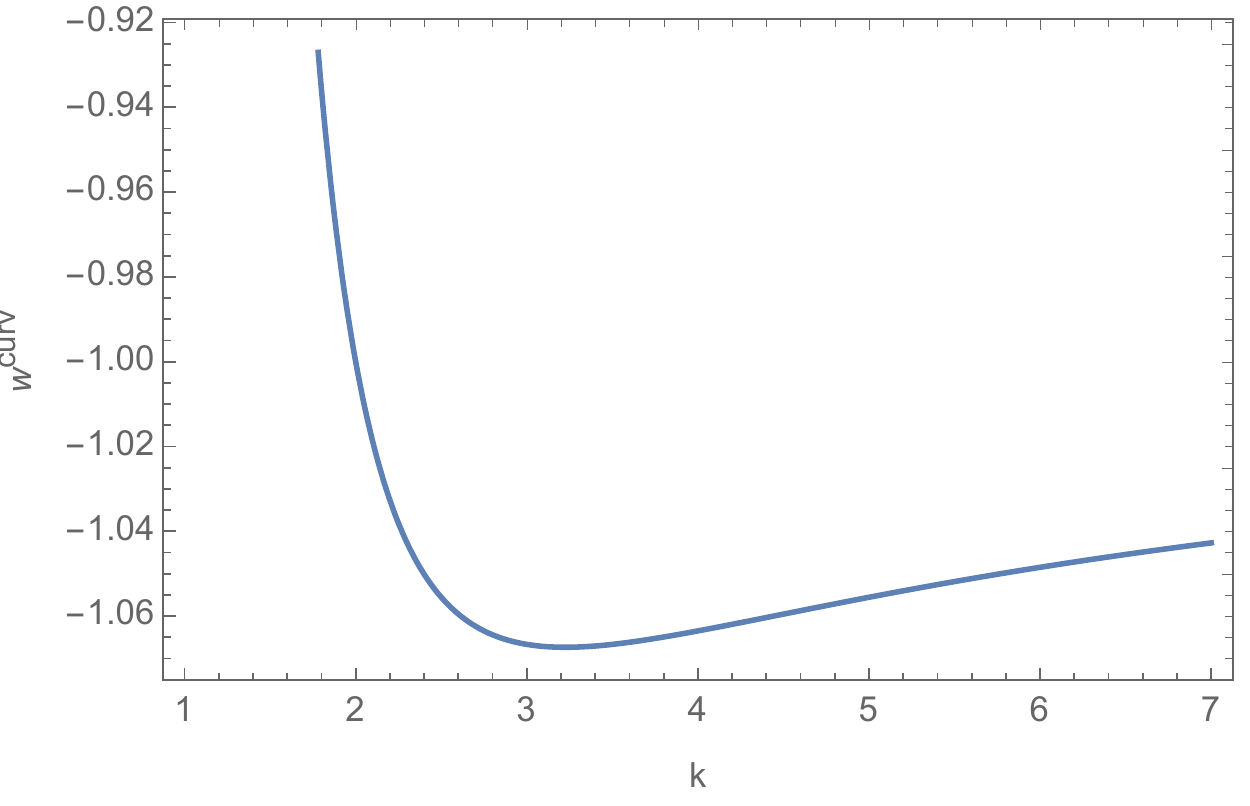}
\caption{Behaviour of the curvature EoS parameter as a function of the parameter $k$.}
\label{fig:EoS}
\end{figure}

The cosmological field equations for $f(R)$ gravity can be recast in terms of effective energy density and pressure given by the extra terms in the gravitational field equations. Specifically, we define 
\begin{align}
&\rho^{curv} = \frac{1}{2}\left(f-R f_R \right)-3 \dot{R} H f_{RR}\,, \label{rho_curv}
\\
&p^{curv} = \frac{1}{2}\left(R f_R-f\right)+(\ddot{R}+2 \dot{R} H) f_{RR}+\dot{R}^{2} f_{RRR}\,. \label{p_curv}
\end{align}
Hereafter, we assume $f_{R}>0$ to have a positive gravitational coupling constant and $f_{RR}>0$ to avoid instabilities \cite{Dolgov:2003px, Faraoni:2006sy}.
In view of the above definitions, the curvature EoS parameter reads
\begin{equation}\label{eq_state}
w^{curv} =
\frac{ -\left(f-R f_R\right)+2(\ddot{R}+2 \dot{R} H) f_{RR}+2\dot{R}^{2} f_{RRR} }{\left(f-R f_R \right)-6 \dot{R} H f_{RR}} \,.
\end{equation}
Taking into account the power-law function emerging from Noether's approach, \emph{i.e.} $f(R) = f_0 R^k$, we obtain 
\begin{equation}
w^{curv} = -\frac{4 k R \dot{a} \dot{R}+a \left[2 k R \ddot{R}+2 k (k-2) \dot{R}^2+R^3\right]}{6
   k R \dot{a} \dot{R}+a R^3}\,.
\end{equation}
Replacing the exponential solution \eqref{expsolf(R)}, we get $w^{curv} =-1$, meaning that this model can potentially match observations on the early inflationary epochs. On the other hand, if we replace the solution \eqref{solf(R)Rk} into Eq.~\eqref{eq_state}, one finds
\begin{equation}
w^{curv} = -\left(\frac{6 k^2-7 k-1}{6 k^2-9 k+3}\right)\,.
\end{equation}
The behaviour of $w$ as a function of $k$ is reported in Fig.~\ref{fig:EoS}.
Notice that dark energy is mimicked when $k$ approaches the value $ k \sim 2$ or for large values of $k$.
Moreover, $w^{curv}$ reaches its minimum at $k \sim 3$, with a small variation with respect to $k\sim 2$, less than $10^{-1}$. 

The dependence of the effective energy density on the scale factor can be found by replacing the solution \eqref{solf(R)Rk} in Eq.~\eqref{rho_curv} and considering the cosmological expression of the Ricci scalar. After some algebra, one obtains
\begin{align}
&\rho^{curv}  =  \frac{f_0 6^k (k-1) (2 k-1) }{2(4 k-5)} \left[\frac{k(k-1)(2 k-1)(4k-5)}{8(k-2)^2}\right]^k \nonumber
\\
&\hspace{1.3cm}\times a^{\frac{2k (k-2)}{(k-1)(2k-1)}}\,.
\end{align}
The energy density $\rho^{curv}$ and the EoS parameter $w$ will be thus used in the next section in order to study the energy conditions and the  inflation in the framework of $f(R)$ gravity. Specifically, $f(R)$ field equations will be recast in terms of effective density $\rho^{curv}$ and pressure $p^{curv}$ with the aim to investigate whether extra geometric terms can behave like exotic fluids.

\subsection{$H_0$ tension and $f(R)$ gravity}
One of the main open problems in modern cosmology is represented by the so-called $H_0$ tension. Specifically, the direct (model-independent) measurement by the Hubble Space Telescope of Cepheids \cite{Riess:2021jrx} and the value of $H_0$ inferred from Planck CMB observations, assuming the $\Lambda$CDM model \cite{Planck:2018jri}, differ from each other by $\sim 5 \sigma$. This evidence has severely put the validity of the $\Lambda$CDM model into question.

However, as confirmed by the most recent Planck findings on the dark energy EoS parameter, \emph{i.e.} $w = -1.041 \pm 0.057$ (at the $68\%$ confidence level)\footnote{The quoted value is reported in the Planck full grid available at \url{http://pla.esac.esa.int/pla/##cosmology}}, allowing for a varying $w$, one can obtain a value for $H_0$ which is slightly higher than the value estimated from the $\Lambda$CDM model \cite{Planck:2018vyg}. This fact may suggest that additional degrees of freedom might be needed in order to cure the $H_0$ problem. This fact is an indication that the issue could be potentially addressed by $f(R)$ gravity. 

For instance, a recent attempt to alleviate the $H_0$ tension is discussed in~\cite{Nojiri:2022ski}, where the authors consider a unified framework where the inflationary era and late-time accelerating era are described in terms of two extra de Sitter vacua.

Another possibility might be to consider a starting gravitational action with various contributions  becoming dominant at different energy scales, meaning that some contributions lead dynamics at early times (whereas the remaining become sub-dominant) and \emph{vice versa} and  others at late times. An example in this direction has been argued in~\cite{Capozziello:2014hia}, where the authors take into account a $f(R)$ gravity of the form
\begin{eqnarray}
        f(R) \simeq && \ldots+\alpha_{(-2)} R^{-2}+\alpha_{(-1)} R^{-1}\nonumber
        \\
        && + \alpha_{(0)} R^{0}+\alpha_{(1)} R+\alpha_{(2)} R^{2}+\ldots
\end{eqnarray}
or, in general
\begin{equation} \label{eq:series}
    f(R) \simeq \sum_{i=-n}^{n} \alpha_{(i)} R^{i},
\end{equation}
with $n \in \mathbb{N}$. These kinds of actions are relevant in string-dilaton gravity in connection to duality properties. See for details \cite{Damour:1994ya, Damour:1994zq}.

Potentially, this model may be able to fit the whole cosmic history from the high energy regimes ($n > 0$) to the large scales ($n<0$), including intermediate scales ($n = 1$), where GR is recovered. The term $\Lambda \equiv \alpha_0 \, R^0$ resembles the case of a pure cosmological constant model. In principle, some contributions of the series \eqref{eq:series} may provide a higher estimate of $H_0$ compared to the $\Lambda$CDM model that may survive at late times and approach the local value provided in \cite{Riess:2021jrx}. A detailed study in this direction will be the argument of a forthcoming paper.

\section{Energy Conditions and Slow-Roll Inflation}
\label{ECSR}
In order to fix the viability of $f(R)$ models, we have to formulate their corresponding energy conditions. These represent a set of inequalities on the energy density and pressure that, if satisfied, suggest whether a model can be physically viable (for example, with respect to causality). Starting from standard matter, their meaning can be generalized to other types of matter, when extra-terms or exotic fluids occur in the gravitational theory. In GR, all the energy conditions are trivially satisfied, unless the cosmological constant is considered. As mentioned in Sec.~\ref{sec:intro}, in some modified theories of gravity, the right-hand side of the field equations can be interpreted as an effective energy-momentum tensor provided by geometry that, in principle, can behave differently with respect to ordinary matter \cite{Capozziello:2013vna, Capozziello:2014bqa}. Here, we show that $f(R)$ gravity does not fulfill the standard energy conditions of GR but violations confirm that our model is capable of properly addressing the accelerating phases of the Universe, mimicking dark energy without introducing any cosmological constant. It is important to stress that, despite possible violations, causality conditions must be respected in any modified gravity model \cite{Capozziello:2013vna, Capozziello:2014bqa}. 

The standard energy conditions of GR are
\begin{itemize}
\item \text{Null Energy Condition (NEC)}: $\rho + p \ge 0$;
\item \text{Weak Energy Condition (WEC)}: $\rho \ge 0 \; ; \; \rho + p \ge 0$;
\item\text{Dominant Energy Condition (DEC)}: $ \rho - |p| \ge 0$;
\item \text{Strong Energy Condition (SEC)}: $\rho + p \ge 0, \, \rho + 3p \ge 0$\,.
\end{itemize}
They are formulated to  select viable states of matter compatible with causality \cite{Hawking:1973uf}. In GR, where the  energy density and pressure are those of  standard fluids, the above system is automatically satisfied. On the other hand, recasting the field equations according to Eqs.~\eqref{rho_curv} and \eqref{p_curv}, effective $\rho^{curv}$ and $p^{curv}$, given by curvature, can be introduced into the energy conditions. In a spatially flat FLRW space-time, the only non-vanishing components of the gravitational field equations are
\begin{subequations}
\begin{align}
& G^0_0 =\frac{1}{f_R}\left[T^0_0 + \frac{1}{2}\left(f-R f_R\right)-3H \dot{R} f_{RR}\right], \\
& G^1_1 = \frac{1}{f_R}\left[T^1_1 +\frac{1}{2}\left(f-R f_R\right) -(\ddot{R}+2 \dot{R} H) f_{RR}-\dot{R}^{2} f_{RRR}\right],
\end{align}
\end{subequations}
with $T^0_0$ and $T^1_1$ being the time and spatial components of the energy-momentum tensor of matter, respectively.
Consequently, assuming a diagonal energy-momentum tensor of the form $T^{\mu}_\nu = \text{diag}(\rho, -p,-p,-p)$, the total energy density $\rho$ and pressure $p$ can be recast in terms of geometry and standard matter as
\begin{align}
& \rho^{tot} =  \rho + \rho^{curv} = \rho +\frac{1}{2}\left(f-R f_R \right)-3 \dot{R} H f_{RR}\,, \\
& p^{tot} = p + p^{curv} = p -\frac{1}{2}\left(f-R f_R\right)+(\ddot{R}+2 \dot{R} H) f_{RR}  \nonumber \\ 
& \qquad \qquad \qquad \qquad \,\,\,\,\, +\dot{R}^{2} f_{RRR}\,.
\end{align}
Hereafter, in order to constrain the free parameters in $f(R)$ gravity, we assume that  matter components satisfy the energy conditions separately. Therefore, introducing the cosmographic parameters $j$ (jerk), $q$ (deceleration) and $s$ (snap) \cite{Weinberg:1972kfs,Capozziello:2017nbu}, we have
\begin{subequations}
\begin{align}
&q = - 1 - \frac{\dot{H}}{H^2},   \;\;\;\; j = 1 + \frac{\ddot{H} + 3 \dot{H} H }{H^3}, 
\\
& s =1 + \frac{\dddot{H} + 3 \dddot{H} H + 3 \dot{H}^2 + 6 H^2 \dot{H} + H \ddot{H}}{H^4} \,,
\label{jqs}
\end{align}
\end{subequations}
so that the scalar curvature and its derivatives become
\begin{align}
&R = - 6 H^2 (1-q), \quad \dot{R} = -6 H^3 (j-q-2)\,,
\\  
&\ddot{R} = -6H^4(s+q^2 +8q+6) \,.
\label{Rqjs}
\end{align}
Thus, replacing the selected model $f(R) \sim R^{k}$ into the energy conditions, we find
\begin{align}
&\rho^{curv} = 18 (k-1) H^4\Big[ - (1-q)^2 + k (j-q-2) \Big]R^{k-2}\,,
\\
&p^{curv} = 36  (k-1) H^6 \Big[-3 (1-q)^3 + k (k-2) (j-q-2)^{2}  \nonumber
\\
&\hspace{1.5cm} + k (s+q^2 +6q+2 + 2j) (1-q) \Big] R^{k-3}.
\end{align} 
Using the numerical values of the cosmographic parameters reported in \cite{Rapetti:2006fv}\footnote{$q = -0.81$, $j = 2.16$ and $s = -0.22$.},  the WEC is violated when $k > -2$ (with $k \neq 1, \, 3)$. Also, the SEC is identically violated for any $k$, so that the deceleration parameter is negative, as expected for an accelerating universe. 

Moreover, further constraints on the values of $k$ may be obtained within the standard inflationary scenario from the so-called slow-roll conditions \cite{Linde:2007fr}:
\begin{equation}
\varepsilon \equiv  \left|-\frac{\dot{H}}{H^2}\right| \ll 1\,, \quad \eta \equiv \left|-\frac{\ddot{H}}{2 H \dot{H}} \right| \ll 1\,.
\label{SRCond}
\end{equation}
In fact, recasting the Friedman equations as
\begin{align}
H^2 & =\frac{1}{3 f_R}\left[\frac{1}{2}\left(f-R f_R\right)-3H \dot{R} f_{RR}\right]\,,
\\
\dot{H} & = - \frac{3}{2} H^2 + \frac{1}{2f_R}\left[\frac{1}{2}\left(f-R f_R\right) \right. 
\\
&\qquad \qquad \qquad \qquad \left.-(\ddot{R}+2 \dot{R} H) f_{RR}-\dot{R}^{2}f_{RRR}\right], \nonumber
\end{align}
and considering the solution \eqref{solf(R)Rk}, one obtains
\begin{equation}
k \ll \frac{1}{2} \left(1-\sqrt{3}\right) \, \lor \, k \gg \frac{1}{2} \left(1+\sqrt{3}\right).
\label{inflrangef(R)}
\end{equation}
These values are consistent with a negative deceleration parameter and, thus, with an accelerated universe.

\section{Swampland Constraints}
\label{SwampSec}

A link between cosmological models and fundamental theories of gravity can  be the swampland criteria \cite{Agrawal:2018own}. Within the string landscape, the conditions over a scalar field potential $V(\phi)$ of a given theory read: 
\begin{itemize}
    \item  the \textit{swampland distance conjecture}, acting on the scalar field and restricting the range of validity of the effective Lagrangian: $\Delta \phi\approx \mathcal{O}(1)$; 
    
    \item the \textit{swampland de Sitter conjecture}, imposing a lower bound condition on
$|\nabla_{\phi}V|/V  \ge \mathcal{O}(1)$. 
\end{itemize}
Swampland criteria have proven to be ruthless with many inflationary models. In particular, it has been shown that they are not satisfied by de Sitter's solutions with a positive cosmological constant \cite{Agrawal:2018own, Garg:2018reu}. 
On the other hand, the analysis on $f(R)$ models, in this context, can be particularly interesting, since models containing symmetries (and, thus, conserved quantities) are invariant under string duality \cite{Benetti:2019smr,Artymowski:2019vfy,Capozziello:2015hra}. 

In order to derive the swampland conditions for $f(R)$ gravity, we can trace the same steps as in Sec.~\ref{studyselectedmodel}. From the $f(R)$ gravity action, one can obtain an effective potential in the Jordan frame as expressed in Eq.~(\ref{potentialconf}), so that the first swampland criterion reads
\begin{equation} \label{eq:S1}
|\Delta \phi| = \bigg|\frac{1}{2} \Delta \ln |f_{R}|\bigg|=\bigg|\frac{1}{2} \frac{1}{f_{R}} f_{RR} \Delta R\bigg| \approx \mathcal{O}(1)\, ,
\end{equation}
where $f_{RR}/f_R$ is the main responsible for satisfying or spoiling this equality.

The second swampland condition can be written as
\begin{align}
\label{eq:S2}
 \frac{|\nabla_{\phi} V|}{V} &=  \bigg|\frac{4}{ R f_{R}-f}\left[ -f +  \frac{R f_{R}}{2} + \frac{f_{R}}{2}\left(\frac{\partial f}{\partial f_{R}}\right)\right. \\
  &\hspace{2.3cm}\left.-\frac{f_{R}^{2}}{2} \left(\frac{\partial R}{\partial f_{R}}\right)\right]\bigg| >  \mathcal{O}(1) \, ,  \nonumber 
\end{align}
which might provide strong constraints on the $f(R)$ models \cite{Benetti:2019smr}. 

In the case of $f(R)\sim R^k$, we obtain 
\begin{equation}
\label{eq: eq_powerlawf_S1}
|\Delta \phi|= \left |\frac{\Delta V}{V} \frac{k-1}{2(2-k)}\right |\lesssim \mathcal{O}(1)\,.  
\end{equation}
where $ \Delta V= \nabla_\phi V \Delta \phi$ and $k \neq 2$. In this case, the constraint on $k$ depends on the ratio $ \Delta V/ V$. Assuming $ \Delta V/ V <1$, Eq.~\eqref{eq: eq_powerlawf_S1} is satisfied for $k<5/3$, while assuming $ \Delta V/ V \lesssim 0.2$ the relation is satisfied for any $k$. 

On the other hand, considering the same form of $f(R)$, in the second swampland criterion we obtain
\begin{equation}
\label{eq:powerlawf_S2}
\frac{|\nabla_{\phi} V|}{V} = \left|\frac{2(2-k)}{k-1}\right| > \mathcal{O}(1)\,, 
\end{equation}
which is defined for $k \neq 1$, and is satisfied for $k<5/3$ and $k>3$. 
As a final consideration, we can infer that for small variations of the potential, \emph{i.e.} $ \Delta V/ V \lesssim 1$,  $k<5/3$ satisfies both swampland conditions. These considerations point out that, in principle, starting from swampland criteria, it is possible to select viable models from the beginning and then reconstruct self-consistent cosmic histories.

\section{Outlook and perspectives}
\label{Concl}
Recent observations are questioning the standard concordance $\Lambda$CDM model as the conclusive paradigm to describe the Universe evolution and dynamics. In addition to the  controversial interpretation of the cosmological constant and the related vacuum energy problem, discrepancies arise also between observations at early and late cosmic times. Motivated by such shortcomings, different modifications to the $\Lambda$CDM model have been proposed in the last years, including theories that extend the gravitational sector. This is the case  of $f(R)$ gravity, which can be considered a first straightforward extension of GR.

Here we applied the Noether Symmetry Approach \cite{Capozziello:1996bi} to select viable models and find the corresponding conserved quantities. Among the  different functions containing symmetries, we focused on the specific model $f(R) \sim R^k$, with $k\in \mathbb{R}$ which allows us to explore possible deviations from GR, which, in turn, is recovered for $k = 1$. We used conserved quantities to reduce  dynamics and find out exact solutions to the Euler-Lagrange equations. Specifically, depending on the value of $k$, it turns out that both time power-law and exponential solutions can be found. The former occur for any $k \neq 1, 1/2, 2$, while the latter for $k = 2$. We also showed that, under conformal transformations, any $f(R)$ model can be cast in the Einstein frame in a dynamically equivalent second-order theory non-minimally coupled to a scalar field. 

Furthermore, we investigated the cosmological dynamics of the selected $f(R)$ model, both in vacuum and in the presence of matter fields. In particular, assuming a homogeneous and isotropic space-time and a universe filled with a perfect fluid, we recast the modified Friedman equations in terms of dimensionless variables, defining an autonomous system of coupled first-order differential equations. Thus, we analyzed the behaviour of the EoS parameter from early times to the current epochs. Specifically, assuming a MD era provided by the standard cosmological scenario, \emph{i.e.} $\Lambda$CDM model, we obtained theoretical bounds on the value of $k$. Results show that the $f(R)$ model under consideration is capable of reproducing the early stages of the Universe if $|k-1|\ll 1$. In vacuum, as shown in Fig.~\ref{fig:EoS}, the EoS parameter resembles the de Sitter behaviour when $k \ge 2$.

We also used the energy conditions to constrain the $f(R)$ theory, proving that different values of $k$ provide models behaving like GR with a cosmological constant. More precisely, the WEC is violated when $k > -2$ ($k \neq 1,3$), while the SEC is identically violated for any $k$. Recasting the field equations in terms of effective energy density and pressure provided by geometry, we studied the slow-roll parameters and showed that standard inflation is allowed by the selected model when $k \ll \frac{1}{2} \left(1-\sqrt{3}\right)$ or $k \gg \frac{1}{2} \left(1+\sqrt{3}\right)$.

Finally, using the potential obtained in the Jordan frame, we checked the validity of the swampland criteria in $f(R) \sim R^k$ gravity. We showed that both criteria are satisfied for $k<5/3$, meaning that $f(R)$ gravity may be helpful also in the ultraviolet regime to overcome issues occurring in the attempts to develop a quantum theory of gravity.

From the above considerations, one may notice that the selected values of $k$ strongly depend on the energy scale under consideration. These values are not always consistent among each other, meaning that no unique $f(R)$ model can explain the whole cosmic history, though, under given limits, the theory is capable of fitting experiments and observations. These disagreements might be due to the fact that different values of $k$ can result dominating at different energy scales. From this point of view, a possible solution might be considering a gravitational action with different terms  which become dominant at certain scales. Of course, a further detailed analysis is needed in order to check whether such a combined action can actually exist and, if so, whether it is capable of fitting all the observations so far available.
The described approach may potentially alleviate the $H_0$ tension which, according to this picture, could be related to different gravity regimes at early and late epochs.

\section*{Acknowledgements}

The authors acknowledge Istituto Nazionale di Fisica Nucleare (INFN), Sezione di Napoli, \textit{iniziative specifiche} GINGER, MOONLIGHT2, QGSKY and TEONGRAV. 
V.D.F. acknowledges Gruppo Nazionale di Fisica Matematica of Istituto Nazionale di Alta Matematica for the support. 

\section*{Data Availability Statement}

Data sharing not applicable to this article as no datasets were generated or analysed during the current study.

\bibliography{references}

\end{document}